\documentstyle[aps]{revtex}
 
\begin{document} 
\title{Geometrical universality in vibrational dynamics}
\author{Raffaella Burioni\footnote{E-mail:
burioni@almite.mi.infn.it}}

\address{Dipartimento di Fisica,
Universit\`a di Milano, via Celoria 16, 20133 Milano, Italy}
\author{Davide Cassi\footnote{E-mail: cassi@vaxpr.pr.infn.it}}
\address{ Dipartimento di Fisica, Universit\`a di Parma,
          Viale delle Scienze, 43100 Parma, Italy 
              }
\date{\today}
\maketitle
\begin{abstract} 
A good generalization of the Euclidean dimension to disordered systems and non
crystalline structures is commonly required to be related to large scale
geometry and it is expected to be independent of local geometrical
modifications. The spectral dimension, defined according
to the low frequency density of vibrational states, appears to be the 
best candidate as far as dynamical and thermodynamical properties are
concerned. In this letter we give the rigorous analytical proof of 
its independence of finite scale geometry. We show that
the spectral dimension is invariant under local rescaling of couplings and 
under addition of finite range couplings, or infinite range couplings decaying 
faster then a characteristic power law. 
We also prove that it is left unchanged by coarse
graining transformations, which are the generalization to graphs and networks
of the usual decimation on regular structures.  
A quite important consequence of all these properties is the possibility of
dealing with simplified geometrical models with nearest-neighbors interactions
to study the critical behavior of systems with geometrical disorder.
   
\end{abstract}
On translationally invariant structures such as crystals, 
many physical phenomena fundamentally depend on the Euclidean dimension $d$,
regardless of the particular form of elementary cells.
This parameter appears to resume all relevant informations about large scale
geometry and therefore it affects the dynamics at low frequencies as well
as the thermodynamics at low temperatures and near critical points. 
In fact this relation is one of the deepest aspects of the physics of
systems with many degrees of freedom. Even if it has not been proved
yet in general, it is commonly accepted and it is experimentally found
that this dimensional dependence is not modified when translational invariance
is lost as a consequence of inhomogeneities or disorder at finite scales.
This invariance under couplings disordering and local changes in 
geometry, which we call {\it geometrical universality}, is perhaps
the most important
feature of dimension in statistical physics. Therefore it is quite reasonable
to require that it is preserved in any meaningful generalization of the 
Euclidean dimension.      

The spectral dimension $\bar d$, firstly introduced by Alexander and Orbach 
\cite{aeo} to characterize the low frequency vibrational spectrum of fractals,
is considered for many reasons the right generalization of $d$ to
non-crystalline structures such as polymers, glasses, fractals, amorphous
materials and so on. In fact $\bar d$ exactly replaces $d$  in most laws where
dimensional dependence explicitly appears: the spectrum of harmonic
oscillations, the average autocorrelation function of random walks, the
critical exponents of the spherical model \cite{sfer}, 
the low temperature specific heat, the generalized Mermin-Wagner theorem
\cite{mwg}, the infrared singularities of the
Gaussian model and many others. In addition $\bar d$ can be easily measured by 
well known experimental techniques such as e.g. neutron scattering.
These remarkable properties suggest extending to non-crystalline
structures the same picture holding for crystals. In other words, we expect
that, even on non translationally invariant networks, the fundamental physical
properties depend only on large scale geometry and that all this dependence is
encoded in $\bar d$.
This would mean that all we need to deal with geometrical disorder is a 
non-integer parameter generalizing the usual dimension and it also would mean
that geometrical disorder does not imply complex behavior.
Clearly, such an appealing picture strongly depends on universal properties of
$\bar d$ which are not obvious neither proved yet. The rigorous proof
of these properties is the subject of this letter.

After introducing the basic concept and notations concerning
$\bar d$, we will prove its geometrical universality following three 
major steps, modifying the networks by acting on the strength of links, 
on the number of links and on the number of points.
First, $\bar d$ is proved to be invariant under a bounded local
rescaling of couplings, which therefore does not alter the underlying 
geometry. Then we will prove its invariance under the addition of 
finite range and infinite range couplings up to a characteristic power law
decay, changing in this way the geometrical structure under consideration
by modifying an infinite number of links.
Finally we will consider a general coarse graining procedure, which changes
the number of degrees of freedom of the network while leaving $\bar d$
invariant. These general transformations allow to group very different 
networks under the same geometrical universality class, characterized by
a given value of $\bar d$. Their significance and applications
will be discussed in details in the conclusions.

The spectral dimension of an infinite connected graph $G$ can be defined 
by studying the low frequency behavior of its vibrational spectrum. 
A graph is a set of points with a topology described by its adjacency matrix
$A_{ij}$, whose elements are equal to $1$ if the points $i$ and $j$ are nearest
neighbors and equal to $0$ in all other cases.

This mathematical structure can be used as a geometrical model for an
oscillating network with masses $m$ at each point, connected along the links 
by springs with elastic constant $k$. Introducing the positive definite Laplacian operator
$L_{ij}\equiv z_i\delta_{ij} - A_{ij}$, $z_i=\sum_j A_{ij}$ being the
coordination number of site $i$, the eigenvalue equations for the vibrational
normal modes read 
\begin{equation}
k\sum_j L_{ij} x_j = \omega^2 m~x_i
\label{oscill}
\end{equation}
where $\omega$ is the frequency and $x_i$ the 
displacement from equilibrium position at site $i$. 
The spectral dimension $\bar d$ is then defined by the
asymptotic behavior of the density of modes with frequency $\omega$, 
$\rho_\omega (\omega)\sim \omega^{{\tilde d}-1}$ for 
$\omega\rightarrow 0$. By (\ref{oscill}), the spectral dimension is 
related to the density $\rho_l (l)$ of eigenstates of the Laplacian with
eigenvalue $l$, whose behavior is given by $\rho_l (l)\sim l^{{\tilde d}/2-1}$
for $l\rightarrow 0$. This relation gives the basis to describe large scale
properties of statistical models in terms of $\bar d$. In particular, we shall
consider the Gaussian model, which provides the most effective tools 
to deal with universal properties of $\bar d$.

The Gaussian model on $G$ is defined by the Hamiltonian:
\begin{equation}
H = {1\over 2}\sum_{ij} \phi_i(L_{ij}+m^2\delta_{ij})\phi_j
\label{ham}
\end{equation}
and its specific free energy $f$ is given by
\begin{equation}
f = -\lim_{N\to \infty} {1\over N} ~F = -\lim_{N\to \infty} {1\over N} \log Z
\end{equation}
where $Z$ is the partition function calculated according to the
Boltzmann weight $\exp(-H)$.
Its correlation functions are given by 
\begin{equation}
<\phi_i\phi_j> = (L+m^2)^{-1}_{ij}
\label{corr}
\end{equation}
and from (\ref{corr}) the matrix $\Phi_{ij} \equiv <\phi_i\phi_j>$ is positive
definite.
The spectral dimension is related to the singular part of $f$ for
$m^2\to 0$ by:
\begin{equation}
Sing~(f) \sim m^{\bar d}.
\label{sing}
\end{equation}
Using this  definition of $\bar d$ and  the above mentioned properties,  a 
``weak" universality of $\bar d$ has been proved, namely its independence of a
local bounded rescaling of masses \cite{debole}, showing that (\ref{sing}) is
left invariant  replacing $m^2$ by $m_i^2=\alpha_i m^2$ in (\ref{ham}),  with
$K^{-1} < \alpha_i < K$ for some $K>1$.

Now our first step is the extension of such invariance to a local bounded
rescaling of couplings. Such ``stronger" universality will provide the
mathematical basis for all further points. 
First of all, we notice that a global rescaling of all couplings by a given
constant $c$ does not affect $\bar d$, since it is equivalent to a global mass
and correlation functions rescaling. 
Then we  prove the invariance of (\ref{sing}) under the rescaling 
$A_{ij}\to J_{ij}=c_{ij} A_{ij}$,
with $K^{-1} < k_{ij} < K$, corresponding to the generalized Laplacian 
$L_{ij}'\equiv z_i'\delta_{ij} - J_{ij}$, $z_i'=\sum_j J_{ij}$.  
The strategy consists in taking successive derivatives of (\ref{sing}) with
respect to $m^2$ up to a divergence for $m^2\to 0$, in proving the monotonicity
of such derivatives with respect to a generic local increasing of the couplings
$J_{ij}$ and then exploiting their boundedness between the two limit cases of
global rescaling $J_{ij}=K^{-1}$ and $J_{ij}=K$ . 
The key point is the monotonicity relation
\begin{equation}
-~\sum_{kl} a_{kl}~ {\partial \over \partial J_{kl} } 
\left(-~ {\partial \over \partial m^2 } \right)^{n} f \ge 0
\label{mono}
\end{equation}
holding for $a_{kl}\ge 0$ and every $n\ge 1$.
This can be proved by noticing that 
\begin{equation}
-~{\partial \over \partial J_{kl} } 
\left(-~ {\partial \over \partial m^2 } \right)^{n} F =
(\Phi^{n})_{kk} + (\Phi^{n})_{ll} - 2 (\Phi^{n})_{kl}
\label{dimostr}
\end{equation}
and that the right hand side of (\ref{dimostr}) is  the value of a positive
definite quadratic form computed on two coinciding vectors $v$, namely 
$\sum_{ij} v_i  (\Phi^{n})_{ij} v_j$, with $v_i=\delta_{ik}-\delta_{il}$.

This first theorem (which we will call {\it strong universality theorem} or
SUT) proves that $\bar d$ depends only on topology and is not
affected by possible disorder in interactions strengths and, together with the
previous result \cite{debole} concerning rescaling of masses, shows that the
low frequency vibrational spectrum is only related to geometry.
In the following, we will see that indeed only {\sl large scale} geometry comes
into play.
The second and third steps deal with transformations affecting the topology
itself, implying a change in the number of links and points.  
It is clear that the changes we are considering must involve sets of links or
points having non vanishing density in the whole graph in the thermodynamic
limit. Indeed, changing a finite number of entry in the Laplacian, or an
infinite one with zero density, corresponds to a compact perturbation and
therefore does not modify its continuous spectrum \cite{kato}.
                                                             
The simplest transformation on links is the addiction of bounded next to
nearest neighbor interactions (NNI). Their exact values are clearly 
irrelevant due to SUT. A possible system $G'$ with such additional interaction 
is described by the generalized Laplacian $L'_{ij}=c L_{ij} -
(L^2)_{ij}$, with $c>2 z_{max}$, which is  
second degree polynomial in $L_{ij}$ lacking of the constant term. Its
eigenvalues $l'$ are related to the eigenvalues $l$ of $L$ by $l'=kl -l^2$ and
they become asymptotically proportional for $l\to 0$. This implies the
asymptotic coincidence of the respective spectral density and, therefore, the
invariance of $\bar d$.  Now, due to SUT, a generic graph obtained from $G$ by
adding some (not necessarily all!) NNI has the same spectral dimension too.
This proof is easily extended to the addition of interactions up to an 
arbitrary finite distance $q$.   
Indeed, iterating $n$ times the same NNI transformation on $G'$, one obtains a
system with interactions to all distances up to $2^{n+1}$. Choosing a suitable
$n$, the invariance of $\bar d$ follows from SUT.
Now, consider the inverse transformation, consisting in acting on $G$ cutting 
bonds which are limited range couplings with respect to the topology of the
resulting graph $G'$. We shall call it {\it bond cutting transformation} (BCT).
From its definition, it follows that BCT preserve $\bar d$. This property will
be very useful in the following.

An analogous strategy can be used to show the invariance of $\bar d$ under 
the introduction of a long range interaction, with power law or
exponential decreasing in the graph chemical distance $r$. The fundamental
observation is that, when the asymptotic
behavior of the eigenvalues $l'$ of the generalized long range Laplacian $L'$
is given by $l'=cl+l^{\alpha}$, with $\alpha > 1$, the new interaction does
not modify $\bar d$, since the above exploited proportionality still holds.

Now, if the long range interaction decreases as $r^{-\gamma}$, where $\gamma >
d_c$ ($d_c$ being connectivity dimension of $G$)
as required by local energy finiteness, it can be shown,
expanding $L'$ on the basis of the eigenvectors of $L$,
that $\alpha={\bar d}(\gamma - d_c)/2 d_c$.          
Therefore $\bar d$ is left unchanged by power law interactions with
$\gamma > d_c + 2 {d_c\over \bar{d}}$ and, from SUT, by any exponential
decaying interaction.
Notice that on Euclidean lattices, where $d_c={\bar d}=d$, this reproduces the
well known result $\gamma > d+2$. 
The addiction of a power law interaction
with a slower decay always alter $\bar d$. From the previous considerations,
$\bar d$ increases according to ${\bar d}'=2 d_c/(\gamma-d_c)$.  

The class of transformations considered up to now does not modify the
number of points of $G$. 

The last kind of graph modifications we will analyze involves a non trivial
change  in the number of degrees of freedom. Actually they include one of the
most important class of transformations used in statistical mechanics, namely
decimations, which are the basic tool of renormalization group on lattices and
fractals.     

We call these transformations {\it topological rescalings}, since they generate
local changes in the number of points. The most general topological rescaling
can be realized through two independent steps. The first one is the {\it
partition} and  consists in dividing 
the graph $G$ in an infinite family of connected subgraphs $G_\alpha$, with
uniformly bounded number of points.  
The second one is the {\it substitution} and consists in generating a new graph
$G'$ by replacing some
or all $G_\alpha$ by a different (connected) graph $S_\alpha$, whose 
number of  points ranges from 1 to a fixed $N_{max}$, and by adding links 
connecting different $S_\alpha$ in such a way that two generic $S_\alpha$ 
and $S_\beta$ are connected by some links if and only if $G_\alpha$
and $G_\beta$ were. 
The simplest topological rescaling occurs when every $S_\alpha$ is composed by
just one point. In this case the resulting graph $G_m$ is called the 
{\it minimal structure} of the partition $\{G_\alpha\}$. The invariance of
$\bar d$ under topological rescaling can be reduced to proving that $G$ and the
minimal structures of its partitions have the same $\bar d$. Indeed, from this
property it easily follows that two graphs having the same minimal structure
also have the same  $\bar d$ and it can be proven that $G$ and $G'$ have the
same minimal structure. We will now show that $\bar d$ is left unchanged
reducing $G$ to its minimal structure. 

For every subgraph $G_\alpha$ let us define the boundary of 
$G_\alpha$, $\partial G_\alpha$, as the set of points of $G_\alpha$ which are 
connected to at least one point of $G_\beta$, $\beta \neq \alpha$. 
We will call $M_\alpha$ the number of points of $\partial G_\alpha$. Let us
consider a new graph $G'$ obtained by replacing each subgraph $G_\alpha$
with the complete graph of its boundary points, $K_{M_\alpha}$
(the complete graph $K_n$ is the $n$-points graph with each point
connected to all the other ones).
It can be shown \cite{hhw} that, considering the Gaussian model 
(\ref{ham}) on $G$ and the Gaussian model on $G'$ with suitable
bounded masses $m'_{i'}$ defined on the points of the $\partial G_\alpha$'s 
and bounded couplings $J'_{i'j'}$ between the points of $K_{M_\alpha}$,
one has for the free energies defined on the two graphs 
$G$ and $G'$:
\begin{equation}
f_{G}=f_{G'}
\label{free}
\end{equation}

The geometry of the new graph $G'$ can be further simplified applying 
the previous transformations.
First, since we are interested in the singular part of $f$,
the exact values of the bounded masses and couplings between the points of
$K_{M_\alpha}$ are irrelevant due to SUT. 

Now consider a single $K_{M_\alpha}$ together with the 
$K_{M_\beta}$
which are connected to it. Then for every $K_{M_\beta}$ select 
one point $i_{\alpha}(\beta)$ among the $M_\alpha(\beta)$ points connected to
$K_{M_\beta}$ and cut the links between all the other 
$M_\alpha(\beta)-1$ points and $K_{M_\beta}$ using BCT. 

Applying again the bonds cutting transformation, 
one can reduce the $i_{\alpha}(\beta)$  
points to a linear chain, observing that all the other links can be
considered as couplings up to k-th neighbors with respect to nearest-neighbors 
interactions between points of the linear chain.

The linear chains can now be shrinked to one point $p_\alpha$
by first applying the previously mentioned  Gaussian model reduction
\cite{hhw},
leading to a new relation for the free energies
as in (\ref{free}), and then cutting the spurious links between external 
points added with this operation using BCT.

Therefore, one is left with a single point $p_\alpha$ 
for each $G_{\alpha}$ connected to the $p_\beta$ corresponding to those
subgraphs $G_{\beta}$ whose points where connected to $G_{\alpha}$ in $G$.
This graph $G_m$ precisely corresponds to the minimal structure of
the initial partition of $G$ and it has the same $\bar d$ as $G$.
 
The three very general classes of geometrical transformations we have been 
considering can be applied in all possible sequences to a graph, leading
to an overall transformation on coupling strength, number of links
and degrees of freedom  which does not change its spectral dimension $\bar d$. 
We will call such a transformation an ${\it isospectrality}$. 
Interestingly enough, isospectralities include a well known class of 
transformations on graphs studied in graph theory, namely the 
rough-isometries \cite{woess}.

Indeed, isospectralities include most part of currently used transformations.
As an example, the usual decimation procedure on fractals 
is a topological rescaling. In particular, 
for all exactly decimable fractals (such as e.g. Sierpinski gaskets and 
T-fractals \cite{amos}), 
the minimal structure of the graph coincides with the 
graph itself. Again, an isospectrality relates the usual two dimensional
square lattice, the hexagonal lattice and the triangular lattice,
which therefore all have dimension $2$.  
In other words, isospectralities are the theoretical formalization of the
intuitive idea of invariance with respect to bounded scale perturbations and
disorder and the {\it isospectrality classes}, defined as the classes of
graphs related by such transformations, are the practical realization of the
apparently abstract concept of non integer dimension.
Now, since most dynamical and thermodynamical properties of generic discrete
structures depend only on $\bar d$, isospectralities provide a very powerful
tool to reduce a very complicated geometrical structure to the simplest one
having the same $\bar d$. The latter  turns out to be much simpler to study and
still presents the same universal properties. 
Moreover, not only an isospectrality can be used to reduce and simplify
structures and problems. It can also be applied, with the opposite aim,  to
build complicated structures with controlled dynamical and thermodynamical
properties, starting from simple deterministic geometrical models. This is the
point of view of {\it spectral dimension engineering}, providing a very
interesting field of possibilities to polymer physicists and material
scientists dealing with non-crystalline materials.

{\bf Figure caption}

Fig.1:  Example of isospectral structures obtained applying isospectral
transformations (without long range couplings) to the T-fractal
(a) and to the  square lattice (b). 
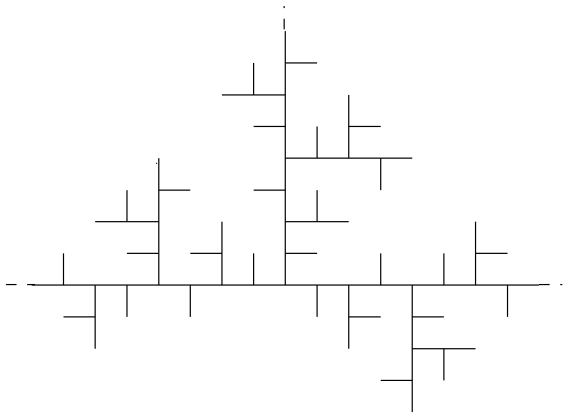
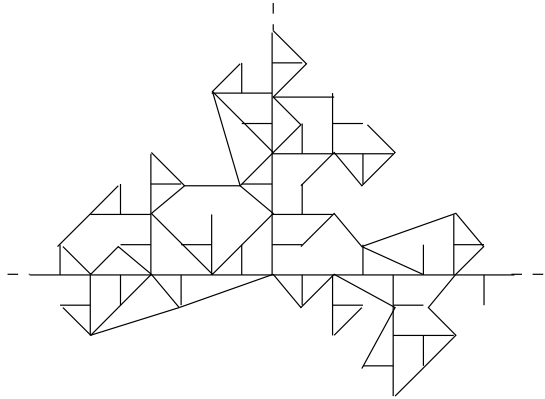
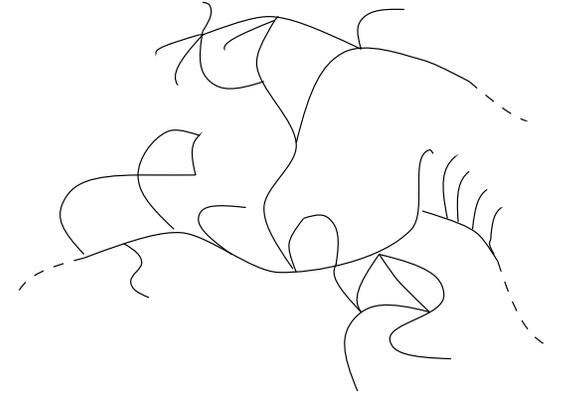
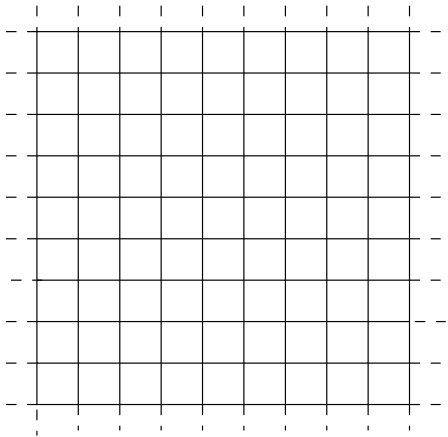
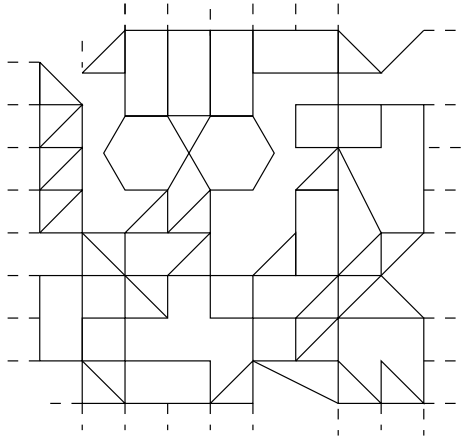
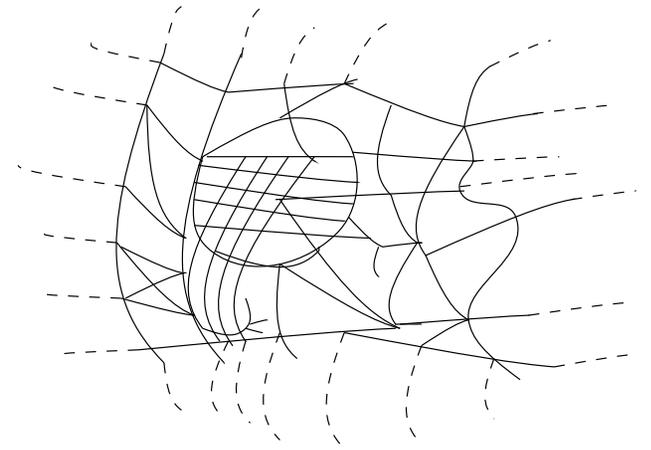
\end{document}